# 人民币汇率变动的"碳减排效应"：来自中国283个地级市的经验证据

**作者信息：**

陈奉先（1982.01-），男，河北玉田，满族，博士研究生学历，首都经济贸易大学金融学院，副教授。主要研究国际资本流动、外汇风险、外汇储备。通讯地址：北京市丰台区花乡张家路口121号首都经济贸易大学（邮编：100070）。通讯方式：（座机）010-83951619，（手机）13683672700，（电邮）cnbanker@163.com，fschan@cueb.edu.cn。

吕潇瑶（1999.11-），女，山西襄汾，汉族，硕士研究生学历，首都经济贸易大学金融学院，研究生在读。主要研究跨国并购与企业生产率。通讯地址：北京市丰台区花乡张家路口121号首都经济贸易大学（邮编：100070）。通讯方式：（手机）15135723089，（电邮）lxy15135723089@163.com。

# 人民币汇率变动的"碳减排效应":来自中国283个地级市的经验证据

**摘要**:本文基于2006~2019年中国283个地级市的面板数据测度了人民币实际有效汇率变动对碳排放强度的影响程度及作用机制。结果表明:(1)人民币实际有效汇率每升值1%,碳排放强度平均下降0.463吨/万元;(2)人民币实际有效汇率升值的"碳减排效应"在东部地区、沿海地区、城镇化水平较高地区、信息公开地区更为明显;(3)人民币实际有效汇率升值通过提升地区的研发创新能力、抑制其外贸和外资水平、促进产业结构优化、改善收入不平等现象,从而降低碳排放强度。

**关键词**:人民币实际有效汇率;对外贸易;产业结构;碳排放强度

**JEL分类号**:F18, F21, O32    **中图分类号**:F293    **文献标识码**:A

## Carbon Emission Reduction Effect of RMB Appreciation: Empirical Evidence from 283 Prefecture-Level Cities of China

**Abstract:** Based on the panel data of 283 prefecture-level cities in China from 2006 to 2019, this paper measures the extent and mechanism of the impact of RMB real effective exchange rate fluctuations on carbon emission intensity. The results show that: (1) For every 1% appreciation of the real effective exchange rate of RMB, the carbon emission intensity decreases by an average of 0.463 tons/10000 yuan; (2) The "carbon emission reduction effect" of RMB real effective exchange rate appreciation is more obvious in the eastern regions, coastal areas, regions with high urbanization levels, and areas with open information; (3) The appreciation of RMB real effective exchange rate can reduce carbon dioxide emission intensity by improving regional R&D and innovation ability, restraining foreign trade and foreign investment, promoting industrial structure optimization and upgrading, and improving income inequality.

# 引言

自 2001 年加入 WTO 以来，中国的土地、劳动力、自然资源等禀赋优势以及出口导向的政策优势得以在更广阔的舞台上充分释放。中国迅速融入国际产业分工并成为"世界工厂"，并在全球价值链中的位置也不断从"低端锁定"到"高端位移"。由此带来中国贸易余额连年增长，从 2001 年的 1865.3 亿元到 2010 年的 12323.3 亿元，再到 2021 年的 43686.7 亿元，二十年间贸易顺差增长约 23 倍。中国外向型发展战略带来经济高速增长的同时也衍生了一系列的问题：资源扭曲问题，生产要素过度集中在出口行业和部门，妨碍其他行业均衡发展；发展脆弱性问题，过于依赖外部导致缺少内生性增长动力、消费驱动的增长模式受阻；更严峻的是环境问题，环境污染、气候变暖、生态系统退化等。

作为世界第二大经济体，全球面临的严峻环境问题决定了我国转变增长发展方式的必要性和必然性。一方面，随着自然资源的逐渐短缺、环境状况的日渐恶化以及劳动力成本的不断提高，我国必须舍弃掉原先高污染、高能耗、高排放的经济发展模式；另一方面，新发展理念为产业升级和绿色发展指明了方向。实现经济的绿色可持续发展不仅是我国生态文明建设的内在要求，也是促进我国经济高质量发展的关键环节。为了改变现状，我国一直秉承人类命运共同体的理念，积极推进碳减排工作。2020 年，习近平总书记在第 75 届联合国大会上明确表明中国在 2030 年前实现"碳达峰"、努力争取在 2060 年前实现"碳中和"的发展目标。党的二十大报告也再次强调，我们要必须要牢固树立和践行绿水青山就是金山银山的理念，积极推进绿色低碳发展，实现人与自然和谐共生。

随着经济全球化的发展和全球价值链分工的深化，汇率日益成为影响一国贸易水平、外汇储备的关键因素，而一个国家的对外贸易不仅会影响到经济领域，还会引起二氧化碳排放的变化。汇率作为连接一国经济领域和二氧化碳排放的一个桥梁，其波动会影响到一国生产要素的积累和配置，从而对该国经济发展方式、产业结构造成一定的影响。近年来，我国一直在积极求变，尝试汇率市场化、增加汇率波动弹性以优胜劣汰。2005 年 7 月 21 日人民币汇率开启市场化改革道路，朝着更具弹性的人民币汇率制度迈进。特别是在 2015 年"8·11"汇改之后，人民币中间价定价机制改革，形成更加透明化、更具市场化的汇率形成机制。在汇率市场化改革的趋势下，人民币兑美元汇率的双向浮动逐渐增强。随着我国对外开放新格局的构建，中国对外开放的程度也会越来越深，汇率波动势必会对国内经济、社会的许多方面产生影响。我国是制造业大国，同时也是世界贸易大国，每年需要从国外进口大量的初级产品来维持正常的生产活动，包括原材料、燃料、中间品等，并将一部分产成品出口到各个国家。汇率波动不可避免地会对工业生产行为、进出口行为产生显著影响，迫使企业及时调整战略以适应新的汇率环境。比如在人民币升值的情况下，处在价值链低端的加工制造业通过产品升级到产业链上层或通过转型进入其他行业以获取更高的利润，最终会导致产业结构发生改变，降低产业发展对资源环境的压力，从而有可能减少地区碳排放。

在此背景下，人民币汇率变动是否会影响地区企业的碳排放？具体作用程度有多大？汇率变动对地区碳排放的作用机制又如何？为了回答这一系列问题，本文首先构造与测算了 2006～2019 年中国地级市层面的人民币实际有效汇率和碳排放强度的数据，考察了人民币汇率变动对各地区碳排放

强度的影响程度以及作用机制。进一步,本文从地区差异、城镇化水平、信息公开程度等角度阐述了人民币汇率变动对地区碳排放强度影响的差异性。本文的结构安排如下:第二章是文献综述及研究假设,第三章是研究设计,第四章是实证分析,第五章是机制分析,第六章是结论与政策建议。

## 文献综述及研究假设

汇率作为一种重要的价格信号,影响着一国参与国际分工的成本,改变着一国生产要素的相对价格,引导着一国资源的有效配置,从而对该国的产业结构、经济增长模式产生重要影响(李向前 et al.,2019;曹伟 et al.,2023),而产业结构和经济增长模式又跟碳排放有着直接的关系。自加入WTO后,我国对外贸易带领国内经济实现了飞速增长。其中,第二产业出口成为我国参与全球生产链和贸易体系的主要方式;与此同时,第二产业出口在我国出口贸易中也占据着绝对领导地位,2019年第二产业出口额占我国出口总额的96.9%。一般而言,出口生产环节伴随着大量的能源投入和消耗,因而以第二产业为主的出口贸易是造成我国碳排放增加的主要"元凶"之一。因此,人民币汇率变动对碳排放的影响可总结如下:第一,当人民币汇率升值时,出口产品丧失了原先的价格优势,短时间内会造成出口量的大幅减少,从而相应地减少能源投入、消耗以及二氧化碳排放(Dogan et al.,2017;Zhang 和 Zhang,2018;Shah et al.,2022)。此外,在市场机制下部分企业会被淘汰,而存活下来的企业在人民币升值背景下会加快技术进步和研发创新力度,提高产品竞争力的同时也会降低企业的碳排放成本,有利于推动企业实现低碳化发展(张兵兵 et al.,2014)。第二,人民币汇率升值可以优化产业间资源配置,促进生产要素从第二产业向第三产业转移,能耗浪费减少,从而有效降低单位产品的碳排放(干杏娣和陈锐,2014;曹伟,2023)。综上,提出本文的第一条假设:

**H1:人民币实际有效汇率升值对地区碳排放具有显著的抑制作用。**

人民币汇率变动是如何影响地区碳排放强度的呢?以往研究认为研发创新、对外开放、产业结构、收入平等是汇率影响地区碳排放强度的重要机制。首先,就研发创新机制而言,汇率波动会通过市场竞争和成本节约两种渠道来影响研发创新行为。第一,市场竞争渠道。当本币升值时,以外币计价的本国产品价格上涨,产品的国际竞争力下降,进而加剧了出口企业间的竞争。企业为了抢占市场会加大研发创新力度、进行技术创新升级、增加产品的附加值、提高产品的科技含量以提升产品的国际竞争优势。因此,当本币升值带来的市场竞争越激烈,企业的创新动机就会越强烈(Ekholm et al.,2012)。第二,成本节约渠道。当本币升值时,企业的进口成本相对降低,企业能够用更少的资金购买到国外的先进技术设备,直接降低企业技术研发成本,进而促使企业不断更新技术,提高管理水平(赵冰和钟霖,2010;康志勇,2015)。

然而,研发创新对碳排放的影响并不确定。一方面,加大研发投入力度、增强自主创新的能力,不仅可以促进新技术的发明创新,而且也有助于增强技术的吸收功能,促使企业有效地学习、吸收国外的先进技术,减少碳排放成本,从而对二氧化碳排放效率的提升起到促进作用(魏巍贤和杨芳,2010);此外,技术进步还可以通过提高能源利用率、优化能源消费结构进而降低碳排放(鄢哲明 et al.,2017;钟超 et al.,2018)。另一方面,科技投入更多地向能促使产出增加、促进经济增长的技术倾斜,而较少地关注绿色低碳技术,对节能减排领域投入力度不够(龚利 et al.,2018);此外,

技术进步在促进经济增长的同时，可能会发生"反弹效应"，导致能源消耗增加，进而使得碳排放增加（Daron *et al.*，2012）。由此，我们提出本文的第二条假设：

**H2a：人民币实际有效汇率升值可能会通过提升地区的研发创新能力，从而降低地区碳排放。**

**H2b：人民币实际有效汇率升值可能会通过提升地区的研发创新能力，从而增加地区碳排放。**

其次，就对外开放机制而言，汇率波动主要会通过对外贸易、外商投资等机制来影响地区的开放度。第一，对外贸易渠道。作为一个出口导向型经济体，中国主要是以制造业产品出口为主，大多数产品是技术含量有限、产品附加值较低，这导致企业的产品定价权弱，产品国际竞争优势主要体现在"物美价廉"上。人民币升值使得出口企业丧失了价格优势，出口规模减少、出口利润受到挤压，也让很多企业在激烈的市场竞争中身陷囹圄（赵冰和钟霖，2010）；低附加值使中国企业出口产品的"依市定价（Price to Market）"能力较弱，人民币汇率变动的影响能够很大程度地传递到出口市场上，从而使企业产品出口量在人民币汇率升值时显著降低（Li *et al.*，2015）。此外既往研究表明，人民币升值降低了进口商品的相对价格，从而有利于先进技术设备的引进以及中间品质量的提高（孙少勤和左香草，2020）。第二，外商投资渠道。大多数学者认为东道国货币升值会通过相对成本效应来抑制 FDI 的流入。一方面，站在外国投资者的角度来看，如果东道国货币升值，外国企业以东道国货币计价的投资成本相对上升，不利于他们在东道国的投资。另一方面，尽管东道国货币升值能降低进口中间投入品的成本，但升值给外商投资者带来的投资成本增加、利润下降的负面影响，可能远超过其带来的进口中间投入品成本下降的正面影响，从而显著降低了外商对华投资的积极性（于燕和杨志远，2014；毛日昇，2015)。

对外开放对碳排放具有双重效应，其影响机理表现在两方面：一方面，对外贸易以及外商投资加强了世界各国之间的联系，一国可以通过引进他国先进的技术、管理经验，对本国企业起到示范作用。在学习过程中，本国企业能够发挥"后发优势"，实现技术和管理水平的赶超，从而优化生产流程，促进能源利用效率提升，最终降低国内的二氧化碳排放（Liang，2014；Sapkota 和 Bastola，2017）；同时，先进的技术经验、严格的环境标准，也促进本国企业碳排放技术的提升，进而提高本国的环境质量（杨传明，2019）。另一方面，随着对外贸易和吸引外资不断增加，我国经济实现了快速增长，但经济快速增长的同时，也会引发一系列的环境问题。在对外贸易的过程中，各国为了保持国际竞争力、抢占市场份额也可能忽略资源的利用效率，导致"向底线赛跑（Race to the Bottom）"[1]的现象，进而增加企业碳排放数量（Esty 和 Dua，1997；Sun *et al.*，2017）。同时，FDI 的引进可能会通过规模效应和污染避难效应来推动碳排放强度提升。FDI 的流入能够推动一个地区经济的快速增长，也会加剧能源、资源的消耗，最终带来碳排放量的增加（Omri *et al.*，2014）；此外，随着发达国家环境规制加强以及环保意识提高，污染密集型产业的环境处理成本较之前明显增

---

[1] "向底线赛跑"的理论认为污染密集型产品出口国，为了保护本国产品在国际市场中的竞争力，倾向于降低本国环境规制水平，并伴随污染密集型企业的地理转移（Kim 和 Wilson，1997；Esty 和 Geradin，2004）。理论上认为"向底线赛跑"的现象仅会出现在环境规制不健全的国家，因为这些国家倾向于屈服于较强的竞争而放弃环境规制，而对于那些环境规制水平较高且已经成熟的国家，环境规制向底线看齐是不会出现的。关于"向底线赛跑假说"是否成立的实证研究也得到了不同的结果。

加，迫使本国将"高污染、高能耗、高排放"的三高企业向环境规制较弱的发展中国家转移，进而增加了发展中国家的二氧化碳排放（路正南和罗雨森，2021）。由此，我们提出本文的第三条假设：

**H3a：人民币实际有效汇率升值可能会通过降低地区的对外贸易水平、外商投资水平，从而降低地区碳排放。**

**H3b：人民币实际有效汇率升值可能会通过降低地区的对外贸易水平、外商投资水平，从而增加地区碳排放。**

然后，就产业结构调整机制而言，汇率波动对产业结构调整有着不同的影响。一方面，我国既是制造业大国又是贸易大国，第二产业出口在我国出口贸易中占据着绝对主导地位，2019年第二产业出口额占我国出口总额的90%以上。人民币升值导致企业产品的出口价格相对提高，而我国出口产品的附加值较低，随着利润空间的下降，企业间优胜劣汰最终会导致第二产业产出减少，倒逼产业结构优化升级（王松奇和徐虔，2015；曹伟，2023）；同时，人民币升值意味着可以用相对较低的价格购入国外先进技术设备，通过先进设备的学习模仿效应和技术溢出效应，研发出具有核心竞争力的相似产品，亦可实现产业结构优化（干杏娣和陈锐，2014；Zhou 和 Ou yang，2018）。

产业结构升级对碳排放具有明显的抑制作用。一般来说，第二产业中高能耗部门占比较高，其产生的二氧化碳也是三类产业中占比最高的。产业结构升级不仅能够降低经济发展对能源和资源的消耗，还促使生产要素在不同产业和行业间进行流转，资源的有效配置有利于提高企业生产效率，加速企业进行绿色转型，进而降低碳排放（孙丽文 *et al.*，2020）；同时，产业结构升级过程中往往伴随着技术创新，技术进步可以通过提高能源利用率、优化能源消费结构、开发新能源从而降低碳排放（黎振强和周秋阳，2021）。由此，我们提出本文的第四条假设：

**H4：人民币实际有效汇率升值可能会通过推动地区产业结构优化升级，从而降低地区碳排放。**

最后，就收入不平等机制而言，汇率波动主要从以下三个机制来影响收入不平等。第一，外商直接投资渠道。当一国货币贬值时，外国企业以东道国货币计价的投资成本相对下降，资本回报率上升，从而吸引外商直接投资流入。外商直接投资增加加大了对劳动力的需求，使得该地区就业率和收入水平的显著提高，导致区域间发展不平衡、收入不平等现象加剧（何枫和徐桂林，2009；马丹和陈紫露，2020）；同时，外商资本的流入主要是加大对技术型人员的需求，扩大了技术性人员和非技术性人员的收入差距，加剧收入不平等现象（任桐瑜和李杰，2021）。另一方面，随着非技术性人员通过培训学习等方式逐渐加入技术性人员的行列中，二者之间的收入差距不断缩小，整个国家的收入不平等现象也得以缓解（谢建国和丁方，2011）。第二，进出口贸易渠道。一国汇率贬值会导致生产资本密集型产品的进口企业资本报酬上升，生产劳动密集型产品的出口企业劳动报酬下降，增加利润率的同时降低工资率，扩大了劳动密集型企业与资本密集型企业之间的收入差距，加剧了收入不平等程度（马丹和陈紫露，2020）。第三，收入分配渠道。一国货币贬值在提高贸易竞争力的同时也给国内提供了更多的就业机会，增加工资份额的同时降低利润份额，从而促进劳动与资本的收入分配，有利于改善收入不平等（李颖和高建刚，2016；马丹和陈紫露，2020）。

然而，收入不平等现象对碳排放的影响并不确定。一方面，消费者生活差异过大会阻碍环保技术的推广，经济欠发达地区的政府在制定政策时可能更倾向于发展经济而忽视环境问题（井波 et al.，2021；张云辉和郝时雨，2022），从而导致碳排放增加。另一方面，与穷人相比富人更加在意环境质量，而环境是一个公共物品，任何人都无法规避，收入不平等在一定程度上可能改善环境质量（Scruggs，1998）。由此，我们提出本文的第五条假设：

**H5a：人民币实际有效汇率升值可能会通过改善地区间收入不平等现象，从而降低地区碳排放。**

**H5b：人民币实际有效汇率升值可能会通过改善地区间收入不平等现象，从而降低地区碳排放。**

# 研究设计

**数据来源** 本文采用地级市层面的数据实证检验人民币实际有效汇率与碳排放强度之间的关系。鉴于数据可获得性，本文选择2006～2019年包括北京、天津、上海、重庆在内的283个地级市面板数据作为研究样本。碳排放数据源自《中国城市统计年鉴》、《中国能源统计年鉴》、万德数据库，人民币实际有效汇率数据源自海关数据库、万德数据库，其他地区层面控制变量数据源自《中国城市统计年鉴》、各省份统计年鉴、地市统计公报、万德数据库。

**估计模型** 基于上述分析，为了检验人民币实际有效汇率对地市碳排放强度的影响，本文设定模型如下：

$$CEI_{it} = a + b\ln REER_{it} + gX_{it} + l_t + h_i + e_{it} \tag{1}$$

其中，$i$表示地级市，$t$表示年份。$REER$表示各地市人民币实际有效汇率，$CEI$表示各地市碳排放强度，$X_{it}$表示控制变量，$l_t$为年份固定效应，$h_i$为个体固定效应，$e_{it}$表示随机扰动项。

**变量设定**

**1. 碳排放强度（CEI）。** 本文采用各地市二氧化碳排放量与实际GDP的比值来衡量。由于无法直接获得各地市二氧化碳排放量的数据，大部分学者采用主要化石能源消费产生的二氧化碳加总来刻画地市级的碳排放数据。本文借鉴韩峰和谢锐（2017）、Glaeser和Kahn（2010）的计算方法，选用天然气、液化石油气以及电力的消费量进行测算，各地市二氧化碳排放量为：

$$CO_2 = C_n + C_p + C_e = kE_n + gE_p + uE_e \tag{2}$$

其中，$C_n$、$C_p$、$C_e$分别表示天然气、液化石油气以及用电产生的二氧化碳排放量。$E_n$表示天然气消费量，$\kappa$表示天然气的二氧化碳排放系数（2.1622kg/m³）；$E_p$表示液化石油气的消费量，$g$表示液化石油气的二氧化碳排放系数（3.1013kg/kg）；$E_e$表示电的消费量，$u$表示各区域电网的二氧化碳排放系数[1]（表1）。中国电网分为华北、华东、华中、东北、西北和南方六大电网，并且公布历年各区域基准线排放因子。

---

[1] 数据来源：中华人民共和国生态环境部.中国电网基准线排放因子
[EB/OL]https://www.mee.gov.cn/ywgz/ydqhbh/wsqtkz/202012/t20201229815386.shtml，2022-11-01

表 1 中国区域电网基准线排放因子

|  | 华北 | 华东 | 华中 | 东北 | 西北 | 南方 |
| --- | --- | --- | --- | --- | --- | --- |
| 2006 | 1.0585 | 0.9411 | 1.2526 | 1.1983 | 1.0329 | 0.9853 |
| 2007 | 1.1208 | 0.9421 | 1.2899 | 1.2404 | 1.1257 | 1.0119 |
| 2008 | 1.1169 | 0.954 | 1.2783 | 1.2561 | 1.1225 | 1.0608 |
| 2009 | 1.0069 | 0.8825 | 1.1255 | 1.1293 | 1.0246 | 0.9987 |
| 2010 | 0.9914 | 0.8592 | 1.0871 | 1.1109 | 0.9947 | 0.9762 |
| 2011 | 0.9803 | 0.8367 | 1.0297 | 1.0852 | 1.0001 | 0.9489 |
| 2012 | 1.0021 | 0.8244 | 0.9944 | 1.0935 | 0.9913 | 0.9344 |
| 2013 | 1.0302 | 0.81 | 0.9779 | 1.112 | 0.972 | 0.9223 |
| 2014 | 1.058 | 0.8095 | 0.9724 | 1.1281 | 0.9578 | 0.9183 |
| 2015 | 1.0416 | 0.8112 | 0.9515 | 1.1291 | 0.9457 | 0.8959 |
| 2016 | 1.0000 | 0.8086 | 0.9229 | 1.1171 | 0.9316 | 0.8676 |
| 2017 | 0.968 | 0.8046 | 0.9014 | 1.1082 | 0.9155 | 0.8367 |
| 2018 | 0.9455 | 0.7937 | 0.8770 | 1.0925 | 0.8984 | 0.8094 |
| 2019 | 0.9419 | 0.7921 | 0.8587 | 1.8026 | 0.8922 | 0.8042 |

最后，我们以2006年为基期，通过GDP指数计算出2006～2019年各地级市的实际GDP，进而计算出各城市对应的碳排放强度（*CEI*）。

**2. 实际有效汇率（*REER*）。** 由于各地市所属的企业都会与不同的国家和地区之间从事贸易业务，各地市的主要贸易伙伴、法定结算币种必然存在明显差异。因此，在同一时期各地市会面临不同的人民币汇率变动情况。为了更加准确地反映各地区受到人民币汇率变动的程度，本文借鉴李宏彬 *et al.*（2011）、余淼杰和王雅琦（2015）计算企业层面实际有效汇率的方法来构建地级市层面的人民币实际有效汇率（*REER*$_{it}$）。地级市$i$在$t$期面对的人民币实际有效汇率（*REER*$_{it}$）如下：

$$REER_{it} = 100 \times \prod_{j=1}^{n} \left( \frac{E_{jt}}{E_{j0}} \times \frac{CPI_{it}}{CPI_{jt}} \right)^{w_{ijt}}, \quad \sum_{j=1}^{n} w_{ijt} = 1 \tag{3}$$

其中，$E_{jt}$表示$t$期$j$国对人民币的汇率，采用间接标价法标价。$E_{j0}$表示基期$j$国对人民币的汇率，本文基期为2006年。$CPI_{it}$表示$t$期地级市$i$的居民消费价格指数（2006年=100），$CPI_{jt}$表示$t$期$j$国家的居民消费价格指数（2006年=100）。$W_{ijt}$表示$i$地级市在第$t$期从$j$国家进出口的货物金额占该地级市同期进出口总金额的比重。按照上述逻辑，我们将海关数据库里的企业按照地级市进行分类，并将地级市内的企业视为一个整体，从而计算出$i$地级市$t$期从$j$国家进出口的货物总额。由于各企业外贸目的地的变更会影响贸易权重，从而影响到地级市的实际有效汇率，造成内生性问题。为了克服此问题，本文将$W_{ijt}$取各个地级市在样本区间内的均值[1]，最终得到*REER*。此外，为了保证量纲一致，在实证分析中对*REER*作对数处理。

为了能够更加直观地反映人民币实际有效汇率与碳排放强度的关系，我们绘制了二者之间的散点图及拟合线，如图1所示：

---

[1] 由于海关数据库目前只公布到2016年，因此我们选用2006～2016年的数据来计算$W_{ijt}$进而求得其均值。

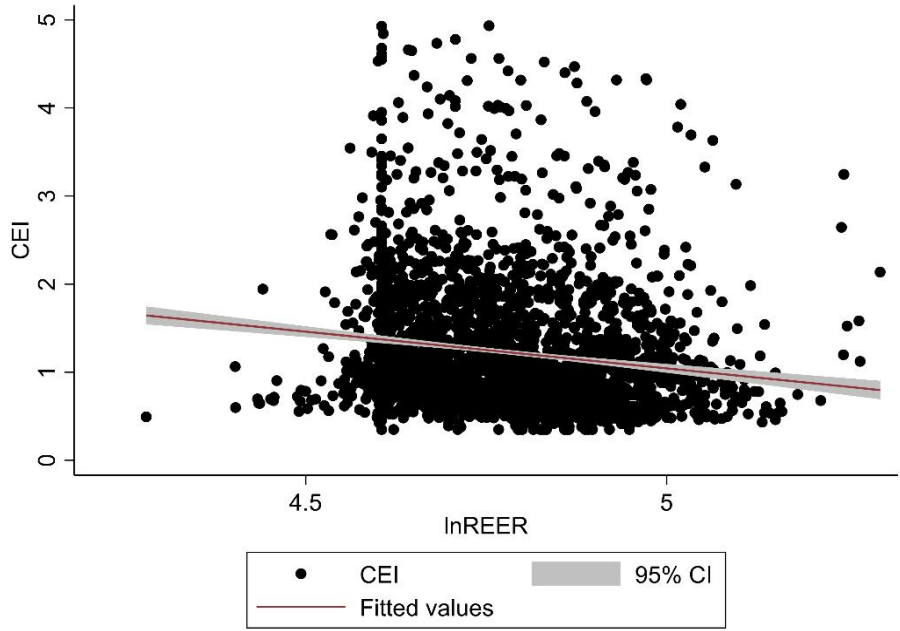

**图 1 人民币实际有效汇率（*lnREER*）与碳排放强度（*CEI*）的散点图与拟合线**

由图1可知，人民币实际有效汇率与碳排放强度之间存在明显的负向关系，至于二者的作用程度及待后续实证检验。

**3. 控制变量**。考虑到影响城市碳排放强度的众多因素，本文借鉴已有的研究成果，引入相关变量，以求控制这些因素干扰，更准确刻画汇率变动对碳排放影响程度，控制变量包括产业结构（*Sec_per*）、人均收入水平（*GDP_per*）及其平方项（*lnGDP_per_sq*）、镇化率（*Urban*）、环境规制（*Env_Regulate*）、森林面积（*lnForestarea*）、碳排放权交易政策（*Policy*）、人力资本（*Human*）、金融发展水平（*Fin_Dvelopment*）、市场化程度（*Market*）。

（1）产业结构（*Sec_per*）

相较于第一和第三产业，第二产业通常是高碳排放产业，其在产业结构中的占比越大，越不利于碳排放的减少。本文采用各地市第二产业增加值占 GDP 的比重来刻画产业结构。

（2）人均收入水平（*GDP_per*）

本文将各城市 2006～2019 年的名义 GDP 进行平减并计算人均实际 GDP，以此刻画人均收入水平。另外根据环境库兹涅茨曲线假说（EKH），人均收入水平和环境质量之间存在着倒 U 型关系，即随着经济的不断增长，环境质量呈现出先恶化而后又改善的态势，因此收入水平与环境质量（碳排放）之间并非简单的线性关系。为了更好地刻画这种非线性关系，本文引入人均收入水平的平方项以考察这种倒 U 型关系。

（3）城镇化率（*Urban*）

中国目前的城镇化已步入了加速发展阶段，"十四五"规划中明确指出要"完善新型城镇化战略、提升城镇化发展质量"。第七次全国人口普查结果显示，截至 2020 年 11 月我国城镇人口比重已达 63.89%。一方面，城镇化意味着城市原有规模的扩大以及新城镇的建设，而城市扩张往往伴随

着大量的基础设施和住宅建设，此过程中会产生大量的二氧化碳排放。另一方面，大量人口从乡村向城镇聚集，原来居分散生产生活方式将会集聚化、中心化并产生的正外部性，从而推动技术创新应用、产业集聚协同和生产效率提升，进而降低环境污染。因此，城镇化率对碳排放强度存在不确定性。本文选用各地市城镇人口占总人口比重来描述城市化水平。

（4）环境规制（*Env_Regulate*）

伴随着我国双碳目标的提出，政府行为在推动绿色经济发展的过程中发挥着重要作用。因此，本文借鉴陈诗一和陈登科（2018）的做法，通过对历年各地市政府工作报告进行文本分析，计算环保相关词汇出现的频率，以此作为地级市层面环境规制指标。

（5）森林面积（*lnForestarea*）

森林在抵消全球碳排放、抑制全球气候变暖中扮演着重要角色。由于缺少地级市森林面积的数据，本文采用各省份森林面积的对数值来衡量一个地区的绿化水平。

（6）碳排放权交易政策（*Policy*）

碳排放权交易市场的建设是我国减少温室气体排放、积极参与全球气候治理、实现"双碳"目标的重要举措，不仅有利于我国绿色低碳经济的发展，也为世界上其他国家兼顾环境治理与经济发展提供可借鉴的经验。一方面，碳排放交易政策的实施通过可以通过引导产业结构调整，促使原有的能源密集型产业向服务业、高科技产业等能源投入少、碳排放量低的产业转型，从而降低能源使用强度，减少环境污染。另一方面，碳排放交易政策的实施可以激发参与主体的绿色创新活力，改进生产技术、提高能源利用效率，从而实现节能减碳。因此，我们引入碳排放权交易政策虚拟变量（Policy）来控制其对各城市碳排放强度的影响。若该城市属于碳排放权交易试点城市且在试点政策实施的当年及之后年份[1]，则 Policy=1，否则 Policy=0。

（7）人力资本（*Human*）

在改善环境质量并实现经济高质量发展的进程中，人力资本是一种不可忽视的重要因素。人力资本水平的提高可以通过提升公众环保意识、促进科技创新来改善环境质量，从而有利于城市实现绿色可持续发展。本文采用各城市本专科生在校学生人数占总人口的比重来衡量各城市的人力资本水平。

（8）金融发展水平（*Fin_Dvelopment*）

一方面，金融发展水平的提高有利于缓解企业进行研发创新时所面临的融资约束，为低碳技术研发提供了资金保障，从而提高企业绿色技术水平、提升资源利用效率、促进产业结构优化升级，进而减少碳排放。另一方面，金融快速发展在一定程度上降低了企业的融资成本，提高了企业扩大生产规模的积极性，从而增加能源消耗、提高碳排放水平。因此，金融发展水平对碳排放的影响尚不确定。本文采用各城市年末金融机构贷款余额占 GDP 的比重来衡量各城市的金融发展水平。

（9）市场化程度（*Market*）

---

[1] 我国于 2013 年正式启动了深圳、北京、天津、广东、上海、湖北、重庆七个碳排放权交易市场。

市场化改革的目的是让市场在资源配置中起决定性作用，提高资源的配置效率。市场化进程的不断推进不仅有助于提高企业的生产资料配置效率，而且打破了地区间壁垒，促进资本、技术等要素的自由流动，进而促使企业学习或引进先进的生产技术以及环保技术，实现低碳发展。本文参照樊纲等（2003）的做法计算出各城市的市场化指数（Market），以此来衡量各城市的市场化程度。

为了剔除极端值对回归结果的影响，本文对所有变量进行了1%的缩尾处理，主要变量的描述性统计如表2所示。

表 2 描述性统计

| 变量 | 符号 | 样本 | 平均值 | 标准差 | 最小值 | P25 | P50 | P75 | 最大值 |
|---|---|---|---|---|---|---|---|---|---|
| 碳排放强度 | CEI | 3962 | 1.33 | 0.94 | 0.35 | 0.73 | 1.04 | 1.61 | 6.00 |
| 实际有效汇率 | lnREER | 2906 | 4.78 | 0.13 | 4.54 | 4.67 | 4.78 | 4.88 | 5.10 |
| 产业结构 | Sec_per | 3962 | 0.48 | 0.11 | 0.20 | 0.41 | 0.48 | 0.55 | 0.76 |
| 人均GDP一次项 | lnGDP_per | 3950 | 1.01 | 0.66 | −0.75 | 0.57 | 1.00 | 1.46 | 2.50 |
| 人均GDP二次项 | lnGDP_per_sq | 3950 | 1.45 | 1.43 | 0.00 | 0.32 | 0.99 | 2.14 | 6.23 |
| 城镇化率 | Urban | 3906 | 0.51 | 0.15 | 0.22 | 0.40 | 0.49 | 0.61 | 0.93 |
| 环境规制 | Env_Regulate | 3880 | 0.01 | 0.00 | 0.00 | 0.00 | 0.00 | 0.01 | 0.01 |
| 森林面积 | lnForestarea | 3962 | 11.05 | 0.81 | 8.54 | 10.55 | 11.18 | 11.52 | 12.47 |
| 人力资本 | Human | 3960 | 0.02 | 0.02 | 0.00 | 0.00 | 0.01 | 0.02 | 0.11 |
| 金融发展水平 | Fin_Dvelopment | 3955 | 0.88 | 0.51 | 0.27 | 0.55 | 0.72 | 1.02 | 2.99 |
| 市场化程度 | Market | 3962 | 10.40 | 2.66 | 4.81 | 8.48 | 10.40 | 12.34 | 16.48 |

## 实证分析

**基准回归** 表1报告了人民币实际有效汇率变动对碳排放强度影响的基准回归结果，其中第（1）列展示的是国家层面的回归结果，我们发现lnREER的系数显著为负且通过1%的显著性水平检验。第（2）—（12)列展示了地级市层面逐步添加控制变量情况下的双固定效应模型结果。我们发现lnREER的系数一直为负且通过5%的显著性水平检验，说明在控制二氧化碳排放强度的时间趋势以及地级市层面不随时间变化的特点之后，人民币汇率变动与碳排放强度呈明显的负相关关系。以第（12）列结果为例，人民币每升值1%，地区的碳排放强度下降0.463吨/万元，从而证明了假设1成立。

表 3 基准回归结果

| | (1) CEI | (2) CEI | (3) CEI | (4) CEI | (5) CEI | (6) CEI | (7) CEI | (8) CEI | (9) CEI | (10) CEI | (11) CEI | (12) CEI |
|---|---|---|---|---|---|---|---|---|---|---|---|---|
| lnREER | **−3.46***** | **−0.425**** | **−0.424**** | **−0.432**** | **−0.429**** | **−0.472**** | **−0.450**** | **−0.447**** | **−0.447**** | **−0.444**** | **−0.441**** | **−0.463**** |
| | (−11.24) | (−2.16) | (−2.18) | (−2.23) | (−2.21) | (−2.47) | (−2.36) | (−2.36) | (−2.36) | (−2.33) | (−2.31) | (−2.38) |
| Urban | | | −1.528*** | −1.420*** | −1.421*** | −1.284*** | −1.178*** | −0.943*** | −0.895*** | −0.897*** | −0.908*** | −0.902*** |
| | | | (−5.31) | (−5.03) | (−5.01) | (−4.64) | (−4.13) | (−3.19) | (−3.01) | (−3.00) | (−3.01) | (−2.97) |
| lnForestarea | | | | −0.478** | −0.466** | −0.449** | −0.452** | −0.424** | −0.417** | −0.404** | −0.419** | −0.404** |
| | | | | (−2.35) | (−2.30) | (−2.22) | (−2.22) | (−2.09) | (−2.06) | (−2.00) | (−2.09) | (−2.02) |
| Human | | | | | −1.766 | −2.238* | −2.270* | −2.131* | −2.359* | −2.318* | −2.327* | −2.502** |
| | | | | | (−1.49) | (−1.82) | (−1.84) | (−1.74) | (−1.91) | (−1.86) | (−1.87) | (−1.97) |
| Fin_Dvelopment | | | | | | 0.142** | 0.152** | 0.128** | 0.129** | 0.128** | 0.126** | 0.111* |
| | | | | | | (2.27) | (2.40) | (2.02) | (2.04) | (2.02) | (2.00) | (1.84) |
| Policy | | | | | | | 0.079** | 0.085** | 0.082** | 0.085** | 0.088** | 0.081** |
| | | | | | | | (2.31) | (2.48) | (2.41) | (2.49) | (2.55) | (2.30) |
| lnGDP_per | | | | | | | | −0.140** | −0.189** | −0.192** | −0.165 | −0.168 |
| | | | | | | | | (−2.00) | (−2.23) | (−2.26) | (−1.49) | (−1.50) |
| lnGDP_per_sq | | | | | | | | | 0.030 | 0.034 | 0.031 | 0.026 |
| | | | | | | | | | (1.17) | (1.31) | (1.06) | (0.90) |
| Market | | | | | | | | | | 0.035 | 0.035 | 0.033 |
| | | | | | | | | | | (1.51) | (1.51) | (1.44) |
| Sec_per | | | | | | | | | | | −0.140 | −0.221 |
| | | | | | | | | | | | (−0.40) | (−0.63) |
| Env_Regulate | | | | | | | | | | | | 2.965 |
| | | | | | | | | | | | | (0.52) |
| Constant | 174.66*** | 3.304*** | 4.113*** | 9.339*** | 9.229*** | 9.053*** | 8.901*** | 8.633*** | 8.532*** | 8.012*** | 8.217*** | 8.213*** |
| | (12.09) | (3.51) | (4.45) | (3.76) | (3.73) | (3.65) | (3.58) | (3.48) | (3.46) | (3.21) | (3.33) | (3.32) |
| 个体固定效应 | 是 | 是 | 是 | 是 | 是 | 是 | 是 | 是 | 是 | 是 | 是 | 是 |
| 时间固定效应 | 是 | 是 | 是 | 是 | 是 | 是 | 是 | 是 | 是 | 是 | 是 | 是 |
| R² | 0.913 | 0.741 | 0.745 | 0.745 | 0.745 | 0.746 | 0.747 | 0.747 | 0.747 | 0.748 | 0.748 | 0.741 |
| 观测值 | 2906 | 2906 | 2864 | 2864 | 2863 | 2861 | 2861 | 2858 | 2858 | 2858 | 2858 | 2817 |

注：括号内为稳健标准误值，***、**、*分别表示在1%、5%和10%的水平下显著，下同。

**稳健性检验**

**1. 替换被解释变量**

碳排放强度是本文的关键变量，为了验证人民币实际有效汇率波动对碳排放强度影响的稳健性，我们对碳排放强度进行了重新测算，选用中国碳核算数据库（*CEADs*）[1]中公布的地区二氧化碳排放量重新计算地区的碳排放强度（*CEI_CEADs*），结果表明 *lnREER* 系数显著为负且通过10%的显著水平检验。该结果与基准回归结果保持一致，说明该结论比较稳健。

**2. 替换解释变量**

人民币实际有效汇率是也是本文的关键变量。为了结果的稳健性，在计算人民币实际有效汇率时，将贸易权重 $W_{ijt}$ 更改为 $i$ 地级市 $t$ 期从 $j$ 国家进口的货物价值占该地级市同期进口总值的比重，再取 $W_{ijt}$ 平均值重新计算人民币实际有效汇率（*REEE_Import*）。我们分别用 *CEI_CEADs*、*lnREEE_Import* 替代基准回归中的 *CEI*、*lnREER* 重新进行回归（见表4）。我们发现 *lnREER*、*lnREER_Import* 系数显著为负且通过5%的显著水平检验，这说明人民币实际有效汇率升值确实能够降低地区的二氧化碳排放强度，与表2基准回归结果保持一致，再次证明了假设1成立。

**3. 加入滞后变量**

考虑到人民币汇率变动对碳排放强度的影响可能存在时滞性，因此我们将 *lnREER* 滞后一期，分别用 *CEI*、*CEI_CEADs* 对其进行回归（见表4），结果表明 *L.lnREER* 系数显著为负且通过10%的显著水平检验。该结果与基准回归仍然保持一致，再次说明该结论比较稳健。

**4. 系统 GMM 检验**

通常而言，系统GMM比差分GMM的估计偏差更小、效率更高，因此本文选用系统GMM法进行估计，回归结果如表4第（6）列所示。由结果可知，*lnREER* 系数显著为负且通过5%的显著水平检验，与预期相符。此外，AR（1）的P值小于0.1，AR（2）的P值大于0.1，这说明扰动项存在一阶自相关但不存在二阶及更高阶的自相关。上述结果表明，本文的核心结论依然稳健。

表4 稳健性检验

|  | （1） | （2） | （3） | （4） | （5） | （6） |
| --- | --- | --- | --- | --- | --- | --- |
|  | CEI_CEADs | CEI_CEADs | CEI_CEADs | CEI | CEI | CEI |
| *lnREER* | **−0.689*** |  |  |  |  | **−0.796**** |
|  | (−1.91) |  |  |  |  | (−2.13) |
| *lnREER_Import* |  | −1.219*** |  | −0.413** |  |  |
|  |  | (−3.21) |  | (−2.33) |  |  |
| *L.lnREER* |  |  | −0.644* |  | −0.302* |  |
|  |  |  | (−1.88) |  | (−1.74) |  |
| *L.CEI* |  |  |  |  |  | 0.763*** |
|  |  |  |  |  |  | (12.39) |

[1] CEADs 团队以化石燃料燃烧以及工业过程产生的 $CO_2$ 排放量来衡量中国地级市的碳排放。其中，化石燃料燃烧产生的 $CO_2$ 排放量是基于 17 种化石燃料和 47 个社会经济部门并采用 IPCC 的方法计算的；工业过程产生的 $CO_2$ 排放量是 9 个工业过程由于化学反应而产生的碳排放。

| | | | | | | |
|---|---|---|---|---|---|---|
| Urban | −0.224 | −0.348 | −0.256 | −0.906*** | −1.188*** | −0.471 |
| | (−0.30) | (−0.46) | (−0.31) | (−2.98) | (−3.57) | (−0.58) |
| lnForestarea | −0.984** | −0.924* | −1.093** | −0.386* | −0.534** | −0.082 |
| | (−2.00) | (−1.90) | (−2.01) | (−1.92) | (−2.22) | (−0.39) |
| Human | −10.159*** | −9.812*** | −8.530*** | −2.431* | −2.001* | 6.176** |
| | (−3.46) | (−3.35) | (−3.05) | (−1.90) | (−1.73) | (2.29) |
| Fin_Dvelopment | 0.099 | 0.100 | 0.088 | 0.108* | 0.113* | 0.075 |
| | (1.06) | (1.06) | (0.93) | (1.78) | (1.85) | (0.67) |
| Policy | 0.262*** | 0.275*** | 0.233*** | 0.092*** | 0.073** | 0.327 |
| | (3.91) | (4.16) | (3.57) | (2.58) | (2.04) | (1.37) |
| lnGDP_per | −0.415* | −0.348 | −0.475* | −0.153 | −0.231* | −0.300 |
| | (−1.85) | (−1.51) | (−1.92) | (−1.37) | (−1.79) | (−1.29) |
| lnGDP_per_sq | 0.294*** | 0.274*** | 0.314*** | 0.022 | 0.042 | 0.023 |
| | (5.34) | (4.85) | (5.18) | (0.78) | (1.21) | (0.37) |
| Market | 0.048 | 0.053 | 0.043 | 0.036 | 0.037 | 0.006 |
| | (1.03) | (1.15) | (0.91) | (1.54) | (1.50) | (0.28) |
| Sec_per | 3.073*** | 3.072*** | 3.157*** | −0.222 | 0.077 | 0.011 |
| | (4.31) | (4.32) | (4.30) | (−0.64) | (0.22) | (0.01) |
| Env_Regulate | −15.449 | −14.926 | −14.225 | 3.005 | 3.625 | 78.707*** |
| | (−1.36) | (−1.33) | (−1.24) | (0.53) | (0.63) | (3.35) |
| Constant | 14.946** | 16.819*** | 15.884** | 7.749*** | 8.851*** | 4.782* |
| | (2.49) | (2.91) | (2.44) | (3.28) | (3.10) | (1.91) |
| 个体固定效应 | 是 | 是 | 是 | 是 | 是 | 是 |
| 时间固定效应 | 是 | 是 | 是 | 是 | 是 | 是 |
| AR(1) | | | | | | 0.001 |
| AR(2) | | | | | | 0.997 |
| $R^2$ | 0.899 | 0.899 | 0.905 | 0.741 | 0.749 | |
| 观测值 | 2323 | 2323 | 2183 | 2817 | 2628 | 2626 |

**异质性分析** 基于上文的研究，我们接下来从地区差异、城镇化水平以及信息公开等角度对样本进行划分，分样本回归结果见表5和表6。

我国经济发展存在着严重的区域不平衡现象，东部沿海地区作为国内最早实施对外开放的区域，是外贸企业最为密集的集聚地，进出口贸易已成为该地区经济发展过程中重要的推动力。相较于中西部、非沿海地区，东部沿海地区在地理位置、自然资源、技术管理等方面拥有良好的条件和优势，再加上改革开放以来东部地区发展迅速，资本较为充裕，为高新技术产业的发展奠定了良好的经济基础。因此，本文将样本中的地级市划分为沿海、非沿海地区以及东部、中西部两组，分组回归结果如表5所示。其中，第（1）-（4）列分别报告的是在沿海、非沿海地区以及东部、中西部样本中，人民币实际有效汇率变动对碳排放强度的回归结果。我们发现，人民币汇率升值的碳减排效果在沿海地区和东部地区显著，而在非沿海地区和中西部地区并不显著。出现这一结果的原因可能是：第一，东部沿海地区交通便利、开放程度高、与全球产业链融合嵌入较深，人民币汇率升值对企业的进出口贸易、外商投资遭受较大冲击，进而对碳排放强度的抑制作用也更为明显。第二，东部沿海

地区主要以出口导向型经济为主，人民币汇率升值后，企业间的竞争加剧，企业为了提升自身产品的竞争力，获得竞争优势，更有动机增加研发投入、提升创新能力，从而为节能减排奠定技术基础。第三，东部沿海地区由于其所具有的先天优势，市场化改革较为彻底，而中西部地区受地理位置影响，市场化改革进程相对迟缓、改革力度不够彻底。一般而言，较高的市场化程度有利于打破市场间壁垒，带来更多的技术交流以及环保技术的扩散，从而使得东部地区碳减排效果更为明显。

表 5 异质性分析—地区差异视角

|  | 沿海 | 非沿海 | 东部 | 中西部 |
| --- | --- | --- | --- | --- |
|  | （1） | （2） | （3） | （4） |
|  | CEI | CEI | CEI | CEI |
| lnREER | **−0.453*** | −0.295 | **−0.531**** | −0.188 |
|  | (−1.80) | (−1.26) | (−2.54) | (−0.72) |
| Urban | 0.849 | −1.039*** | 0.495 | −1.472*** |
|  | (1.07) | (−3.15) | (1.05) | (−3.89) |
| lnForestarea | −0.598** | −0.192 | −0.046 | −0.990** |
|  | (−2.05) | (−0.80) | (−0.29) | (−2.12) |
| Human | −2.744** | −2.060 | −1.440 | 1.154 |
|  | (−2.08) | (−1.30) | (−1.13) | (0.53) |
| Fin_Dvelopment | 0.085 | 0.136* | 0.011 | 0.128 |
|  | (1.19) | (1.81) | (0.20) | (1.53) |
| Policy | −0.083 | 0.100** | 0.041 | 0.087 |
|  | (−1.36) | (2.20) | (0.90) | (1.22) |
| lnGDP_per | 0.313 | −0.061 | 0.131 | −0.047 |
|  | (1.56) | (−0.46) | (1.01) | (−0.29) |
| lnGDP_per_sq | −0.109** | −0.019 | −0.026 | −0.090* |
|  | (−2.35) | (−0.55) | (−0.75) | (−1.95) |
| Market | 0.025 | 0.033 | −0.008 | 0.054 |
|  | (0.94) | (1.18) | (−0.36) | (1.57) |
| Sec_per | −0.743 | −0.305 | −0.241 | 0.255 |
|  | (−1.18) | (−0.73) | (−0.59) | (0.49) |
| Env_Regulate | −17.535* | 12.727* | −13.719* | 7.834 |
|  | (−1.71) | (1.91) | (−1.91) | (0.96) |
| Constant | 9.143*** | 5.110* | 3.956* | 13.368** |
|  | (2.62) | (1.73) | (1.88) | (2.41) |
| 个体固定效应 | 是 | 是 | 是 | 是 |
| 时间固定效应 | 是 | 是 | 是 | 是 |
| $R^2$ | 0.705 | 0.750 | 0.685 | 0.756 |
| 观测值 | 637 | 2180 | 1120 | 1697 |

城镇化率是衡量地区发展水平的重要指标，城镇化水平的提高不仅意味着大量人口从农村涌入城镇，还意味着人们生活方式、经济发展模式等多方面的变化。在深入推进城镇化过程中，一方面较高的城镇化率会加大对基础设施和住宅的建设需求，从而导致城市能源消费、碳排放增加；另一方面，分散生产生活方式将会集聚化、中心化并产生的正外部性，从而推动技术创新应用、产业集聚协同和生产效率提升，进而降低环境污染。因此，本文按照城镇化率（Urban）的中位数将样本中的地级市划分为两组，分组回归结果如表6所示。其中，第（1）、（2）列分别报告的是在城镇化水平高、低两组样本中，人民币实际有效汇率变动对碳排放强度的回归结果。结果表明人民升值对显

著抑制了城镇化水平较高城市的碳排放强度，而对城镇化水平较低城市并没有产生显著作用。出现这一结果可能的原因是：与城镇化水平较低的城市相比，城镇化水平较高的城市集聚效应和规模效应更明显，从而使得该城市内各企业的能源利用效率、技术水平显著提高，进而节能减碳的效果也更加显著。

2008年公众环境研究中心和自然资源保护委员会开发并逐年公布了"污染源监管信息公开指数"的113城市，2013年增加了镇江、三门峡、自贡、南充、德阳、玉溪、渭南7个城市。因此，本文以PITI指数中涉及的120个重点环保城市为准，按照是否属于该120个城市将样本中的地级市划分为两组，回归结果如表6所示。其中，第（3）、（4）列分别报告的是在信息公开、非信息公开两组样本中，人民币实际有效汇率变动对碳排放强度的回归结果。结果表明人民升值对显著抑制了信息公开城市的碳排放强度，而对非信息公开城市城市并没有产生显著作用。出现这一结果可能的原因是：一般情况下，信息公开主要是针对环保重点城市，其会倒逼该城市通过推动产业结构升级、技术创新等来达到保护环境的效果。因此，由于受环境信息公开制度的影响，环境信息公开的城市在面临人民币汇率升值时，提升创新能力的动力更强，从而使得地区的碳减排效果更为明显。

表6 异质性分析—城镇化水平、信息公开视角

|  | 高城镇化 | 低城镇化 | 信息公开 | 非信息公开 |
|---|---|---|---|---|
|  | （1） | （2） | （3） | （4） |
|  | $CEI$ | $CEI$ | $CEI$ | $CEI$ |
| $lnREER$ | **−0.876***** | −0.064 | **−0.666***** | −0.285 |
|  | (−3.49) | (−0.22) | (−2.63) | (−1.09) |
| $Urban$ | 0.064 | −2.604*** | 0.407 | −1.584*** |
|  | (0.17) | (−4.20) | (0.89) | (−3.74) |
| $lnForestarea$ | −0.819*** | −0.845** | −0.573*** | 0.051 |
|  | (−2.98) | (−2.12) | (−2.62) | (0.13) |
| $Human$ | −1.138 | 3.950 | −2.344* | 1.405 |
|  | (−0.89) | (0.53) | (−1.66) | (0.48) |
| $Fin\_Dvelopment$ | 0.107 | −0.209* | 0.181*** | 0.082 |
|  | (1.55) | (−1.69) | (2.61) | (0.73) |
| $Policy$ | −0.017 | 0.351*** | 0.031 | 0.134** |
|  | (−0.38) | (4.84) | (0.78) | (2.32) |
| $lnGDP\_per$ | −1.197*** | 0.273 | −0.786*** | 0.072 |
|  | (−4.96) | (1.36) | (−5.18) | (0.38) |
| $lnGDP\_per\_sq$ | 0.292*** | −0.363*** | 0.215*** | −0.031 |
|  | (4.29) | (−5.05) | (5.41) | (−0.65) |
| $Market$ | −0.002 | 0.022 | 0.032 | 0.009 |
|  | (−0.09) | (0.60) | (0.95) | (0.28) |
| $Sec\_per$ | 0.616 | −1.600*** | 0.030 | −0.449 |
|  | (1.31) | (−2.75) | (0.06) | (−0.87) |
| $Env\_Regulate$ | 2.548 | 25.002** | −2.451 | 9.151 |
|  | (0.39) | (2.37) | (−0.31) | (1.03) |
| $Constant$ | 14.871*** | 12.774*** | 10.504*** | 2.858 |
|  | (4.13) | (2.76) | (3.58) | (0.65) |
| 个体固定效应 | 是 | 是 | 是 | 是 |
| 时间固定效应 | 是 | 是 | 是 | 是 |
| R² | 0.808 | 0.766 | 0.788 | 0.689 |
| 观测值 | 1579 | 1227 | 1425 | 1392 |

# 机制分析

首先，一国货币升值会促进本国企业增加创新投入、引进先进技术，而技术进步一方面可能会通过提高能源利用率、优化能源消费结构降低碳排放强度，另一方面可能更多地应用到提高产能上，而较少的关注低碳技术。其次，一国货币升值会在不同程度上抑制企业的进出口贸易、外商直接投资，而外对贸易和外商来华投资的减少一方面降低企业对资源的消耗、避免成为发达国家的"污染天堂"，另一方面削弱了与世界各国在生产和环境技术、管理经验以及环境标准等方面的交流。然后，一国货币升值一方面抑制劳动密集企业的出口、促进先进技术的引入实现产业结构优化升级，另一方面会给外资企业带来良机的同时对国内高新技术产业发展造成了负向冲击，而产业结构优化升级会降低对资源和能源消耗、促进生产要素的有效配置、激发创新活力从而加速企业绿色转型。最后，一国货币升值会通过收入分配等路径来改善收入不平等现象，从而有效改善环境质量。下面，我们实证检验研发创新能力、贸易投资、产业结构、收入不平等现象这四种机制是否成立。

**研发创新机制** 由于《中国城市统计年鉴》缺乏地级市层面的研发投入经费数据，我们借鉴余泳泽 et al.（2020）的做法，采用各地市科研综合技术从业人员占从业人员的比重来衡量地级市的研发投入强度（$R\&D\_Intensity$）。同时为了提高结果的稳健性，我们选用每十万人中绿色发明专利的授权数来衡量地级市的绿色创新能力（$Green\_innovation$）。我们按照 $R\&D\_Intensity$、$Green\_innovation$ 的中位数将样本中的地级市划分为两组，回归结果如表 7 所示。结果表明，在研发投入和绿色创新低组中 lnREER 系数在5%的水平上显著为负，而在研发投入和绿色创新高组系数并不显著。我们认为人民币升值推动创新水平较低的城市大幅提升创其新能力尤其是绿色创新能力，从而使得该类城市因技术水平提升而给环境带来的边际贡献更为突出，对碳排放强度的抑制作用也更为明显。因此，人民币升值的碳减排效应在创新能力低的城市发挥了更大的作用，从而证实了假设 H2a 的存在。

表 7 研发创新渠道检验

| | 低研发投入 | 高研发投入 | 低绿色创新 | 高绿色创新 |
|---|---|---|---|---|
| | （1） | （2） | （3） | （4） |
| | CEI | CEI | CEI | CEI |
| lnREER | **−0.600**** | −0.142 | **−0.505**** | −0.648 |
| | (−2.20) | (−0.49) | (−2.30) | (−1.62) |
| Urban | −1.201** | −0.023 | −0.514 | −1.659*** |
| | (−2.39) | (−0.06) | (−1.06) | (−3.21) |
| lnForestarea | −0.467 | −0.124 | 0.021 | −0.874*** |
| | (−1.64) | (−0.43) | (0.06) | (−3.19) |
| Human | −8.716*** | −1.321 | −1.333 | −2.704** |
| | (−2.64) | (−0.99) | (−0.36) | (−2.00) |
| Fin_Dvelopment | 0.166 | 0.067 | 0.039 | −0.021 |
| | (0.97) | (1.08) | (0.43) | (−0.27) |
| Policy | 0.218*** | −0.023 | 0.239*** | −0.054 |
| | (3.84) | (−0.51) | (2.96) | (−1.32) |
| lnGDP_per | 0.070 | −0.239 | −0.142 | −0.027 |

|  | (0.41) | (−1.36) | (−0.81) | (−0.10) |
| --- | --- | --- | --- | --- |
| lnGDP_per_sq | 0.057 | −0.022 | 0.007 | −0.078 |
|  | (1.30) | (−0.47) | (0.12) | (−0.96) |
| Market | 0.137*** | −0.079*** | 0.075* | −0.041 |
|  | (3.36) | (−2.88) | (1.85) | (−1.40) |
| Sec_per | −0.297 | −0.102 | −0.345 | −0.244 |
|  | (−0.54) | (−0.20) | (−0.67) | (−0.42) |
| Env_Regulate | −1.607 | 3.430 | 9.465 | 3.270 |
|  | (−0.17) | (0.52) | (0.94) | (0.49) |
| Constant | 8.514** | 4.379 | 3.330 | 15.613*** |
|  | (2.51) | (1.16) | (0.83) | (4.30) |
| 个体固定效应 | 是 | 是 | 是 | 是 |
| 时间固定效应 | 是 | 是 | 是 | 是 |
| $R^2$ | 0.756 | 0.765 | 0.817 | 0.764 |
| 观测值 | 1383 | 1402 | 1199 | 1540 |

**对外开放机制** 首先，本文采用各地市进出口总额占 GDP 的比重（Trade_GDP）来衡量城市的对外贸易水平，同时为了探究到底是进口贸易还是出口贸易在碳减排过程中发挥作用，又分别用各地市进口总额占 GDP 的比重（Import_GDP）、出口总额占 GDP 的比重（Export_GDP）来衡量进出口水平。我们按照 Trade_GDP、Import_GDP、Export_GDP 的中位数将样本中的地级市划分为两组，回归结果如表 8 所示。结果表明在高对外贸易、高进口、高出口组中 lnREER 系数均在 1%的水平上显著为负，而低高对外贸易、高进口、高出口组中该系数并不显著。出现这一结果可能的原因是：第一，就出口水平而言，人民币实际有效汇率上升时，我国的出口商品丧失了原先的价格优势，不利于我国企业进行出口贸易，从而有效减少了资源密集型产业对能源和资源的浪费，因此，人民币汇率升值的碳减排效应在出口水平较高的城市发挥了更大的作用。第二，就进口水平而言，人民币升值可以降低企业进口中间品的相对价格，激励企业进口更高质量的中间品，从而产生技术溢出效应，提高 TFP 的同时也有助于推动环境质量的改善，因此，人民币汇率升值的碳减排效应在进口水平较高的城市发挥了更大的作用。

然后，本文各地市外商实际投资额占 GDP 的比重（Fdi_GDP）来衡量城市的外商来华投资水平。我们按照 Fdi_GDP 的中位数将样本中的地级市划分为两组，回归结果如表 8 所示。结果表明在高外商投资组中 lnREER 系数在 1%的水平上显著为负，而低外商投资组中该系数并不显著。这也就是说，相对于低外商投资城市而言，人民币实际有效汇率升值对碳排放强度的负向影响在高外商投资城市中发挥着更加突出的作用。事实上，人民币升值使得外商投资者手中的财富变得相对较少，来华投资的成本变高，从而抑制了 FDI 的流入，也使我国避免成为他国的"污染天堂"。因此，人民币汇率升值的碳减排效应在的高外商投资水平的城市中发挥了更明显的作用，综上证实了假设 H3a 的存在。

表 8 对外开放机制检验

|  | 高对外贸易 | 低对外贸易 | 高进口 | 低进口 | 高出口 | 低出口 | 高外商投资 | 低外商投资 |
| --- | --- | --- | --- | --- | --- | --- | --- | --- |
|  | （1） | （2） | （3） | （4） | （5） | （6） | （7） | （8） |
|  | CEI | CEI | CEI | CEI | CEI | CEI | CEI | CEI |
| lnREER | −1.007*** | 0.252 | −0.752*** | −0.035 | −0.912*** | 0.211 | −0.896*** | 0.134 |

| | (−4.96) | (0.76) | (−3.00) | (−0.13) | (−4.06) | (0.63) | (−4.21) | (0.44) |
|---|---|---|---|---|---|---|---|---|
| Urban | −0.351 | −2.224*** | −0.234 | −1.796*** | −0.562 | −1.479*** | −0.534 | −0.656 |
| | (−0.95) | (−3.92) | (−0.69) | (−3.54) | (−1.19) | (−3.73) | (−1.59) | (−0.88) |
| lnForestarea | −0.196 | −0.442 | −0.361* | −0.281 | −0.258 | −0.929* | −0.205 | −0.428 |
| | (−1.01) | (−0.90) | (−1.71) | (−0.67) | (−1.30) | (−1.89) | (−1.17) | (−0.67) |
| Human | −3.317** | 3.630 | −1.745 | −0.829 | −4.845*** | 4.474 | −2.950** | −5.284 |
| | (−2.20) | (1.10) | (−1.41) | (−0.20) | (−2.92) | (1.37) | (−2.11) | (−1.02) |
| Fin_Dvelopment | 0.044 | 0.006 | 0.060 | 0.137 | 0.032 | −0.006 | 0.172*** | −0.037 |
| | (0.69) | (0.05) | (0.78) | (1.47) | (0.51) | (−0.05) | (3.07) | (−0.29) |
| Policy | 0.012 | 0.180** | 0.012 | 0.107 | −0.007 | 0.206*** | 0.045 | 0.182** |
| | (0.31) | (2.43) | (0.29) | (1.55) | (−0.16) | (3.02) | (1.28) | (2.13) |
| lnGDP_per | −0.391*** | 0.164 | −0.678*** | 0.249 | −0.295* | −0.011 | −0.283** | 0.026 |
| | (−2.89) | (0.80) | (−4.66) | (1.28) | (−1.91) | (−0.06) | (−2.08) | (0.13) |
| lnGDP_per_sq | 0.115*** | −0.178*** | 0.169*** | −0.132** | 0.088** | −0.147** | 0.117*** | −0.048 |
| | (3.22) | (−2.67) | (4.59) | (−2.37) | (2.35) | (−2.57) | (3.41) | (−0.77) |
| Market | −0.018 | 0.083* | −0.024 | 0.083** | −0.041* | 0.094** | 0.025 | 0.048 |
| | (−0.83) | (1.69) | (−1.06) | (2.09) | (−1.78) | (2.20) | (0.91) | (1.13) |
| Sec_per | 0.003 | −1.121* | −0.016 | −1.206** | −0.451 | −0.726 | 0.215 | −0.856 |
| | (0.01) | (−1.85) | (−0.03) | (−2.23) | (−1.00) | (−1.36) | (0.51) | (−1.44) |
| Env_Regulate | −8.529 | 17.672 | −1.193 | 23.608** | −9.658 | 13.132 | −3.693 | 12.834 |
| | (−1.34) | (1.55) | (−0.18) | (2.50) | (−1.49) | (1.19) | (−0.54) | (1.37) |
| Constant | 8.923*** | 5.610 | 9.693*** | 4.847 | 9.696*** | 10.813* | 7.626*** | 5.762 |
| | (3.43) | (1.00) | (3.49) | (1.01) | (3.59) | (1.91) | (3.58) | (0.76) |
| 个体固定效应 | 是 | 是 | 是 | 是 | 是 | 是 | 是 | 是 |
| 时间固定效应 | 是 | 是 | 是 | 是 | 是 | 是 | 是 | 是 |
| $R^2$ | 0.827 | 0.683 | 0.816 | 0.750 | 0.813 | 0.746 | 0.736 | 0.757 |
| 观测值 | 1706 | 1092 | 1635 | 1117 | 1689 | 1058 | 1556 | 1174 |

**产业结构升级机制** 本文采用各地市第二产业产值占 GDP 的比重（*Sec_per*）来衡量城市的产业结构。我们按照 *Sec_per* 的中位数将样本中的地级市划分为两组，回归结果如表 9 所示。其中，第（1）、（2）列分别报告的是在高、低水平组中，人民币实际有效汇率变动对碳排放强度的回归结果。结果表明在第二产业占比高组中 *lnREER* 系数在 1%的水平上显著为负，而在第二产业占比低组中 *lnREER* 系数不显著。这也说明人民币升值在第二产业占比较高的城市中起到了优化产业结构的效果，进而实现碳减排。出现这一结果可能的原因是：人民币升值时，第二产业占比较高的城市中加工出口贸易受到的负向冲击更为明显，倒逼企业转型升级，有效减少了第二产业对能源和资源的浪费，从而显著抑制了碳排放。因此，人民币升值的碳减排效应在第二产业占比较高的城市中发挥了更大的作用，从而证实了假设 H4a 的存在。

**收入不平等机制** 本文借鉴王少平和欧阳志刚（2008）的研究，采用包含城乡收入和人口的泰尔指数（*Theil*）作为收入不平等的衡量指标。我们按照泰尔指数（*Theil*）的中位数将样本中的地级市划分为两组，回归结果如表 9 所示。其中，第（3）、（4）列分别报告的是在收入不平等现象强、弱水平组中，人民币实际有效汇率变动对碳排放强度的回归结果。结果表明在收入不平等现象较强的组中 *lnREER* 系数在 1%的水平上显著为负，而收入不平等现象较弱中 *lnREER* 系数不显著。这也说明人民币升值在收入不平等现象较强城市中起到了更明显的碳减排作用。出现这一结果可能的原因是：人民币升值使得外商直接投资、劳动密集产品出口明显减少，避免出现不同行业收入差距过

大，改善收入不平等现象。因此，越是收入差距过大的城市，受到人民币升值的影响就越大，越有利于改善收入不平等，从而证实了假设 H5a 的存在。

表 9 产业结构升级、收入不平等机制检验

|  | 第二产业占比高 | 第二产业占比低 | 收入不平等现象较强 | 收入不平等现象较弱 |
|---|---|---|---|---|
|  | （1） | （2） | （3） | （4） |
|  | CEI | CEI | CEI | CEI |
| lnREER | **−1.000***** | 0.038 | **−1.424**** | 0.442 |
|  | (−3.33) | (0.17) | (−3.18) | (0.76) |
| Urban | −0.536 | −0.684 | −0.777 | 2.478** |
|  | (−1.30) | (−1.28) | (−0.65) | (2.25) |
| lnForestarea | −0.563** | 0.245 | −3.387** | −0.382 |
|  | (−2.16) | (0.75) | (−2.31) | (−0.84) |
| Human | −5.327** | −1.890 | −21.335** | −6.739*** |
|  | (−2.11) | (−1.30) | (−2.15) | (−2.61) |
| Fin_Dvelopment | 0.075 | 0.016 | 0.030 | −0.025 |
|  | (0.51) | (0.24) | (0.19) | (−0.20) |
| Policy | 0.072 | 0.097* | 0.179 | 0.466*** |
|  | (1.01) | (1.91) | (1.02) | (5.21) |
| lnGDP_per | −0.613*** | −0.085 | 0.706* | −1.199*** |
|  | (−2.87) | (−0.56) | (1.90) | (−3.46) |
| lnGDP_per_sq | 0.215*** | −0.025 | −0.007 | 0.468*** |
|  | (3.38) | (−0.70) | (−0.08) | (4.81) |
| Market | 0.130*** | −0.012 | −0.087 | 0.073 |
|  | (3.03) | (−0.51) | (−0.94) | (1.19) |
| Sec_per | 0.395 | −1.293** | 3.777*** | 1.717* |
|  | (0.56) | (−2.56) | (3.25) | (1.70) |
| Env_Regulate | 4.776 | 2.274 | −20.213 | 11.023 |
|  | (0.50) | (0.36) | (−0.95) | (0.86) |
| Constant | 11.406*** | −0.605 | 46.383*** | 1.873 |
|  | (3.36) | (−0.17) | (2.82) | (0.31) |
| 个体固定效应 | 是 | 是 | 是 | 是 |
| 时间固定效应 | 是 | 是 | 是 | 是 |
| $R^2$ | 0.797 | 0.739 | 0.921 | 0.929 |
| 观测值 | 1476 | 1317 | 967 | 1325 |

## 结论与政策建议

本文首先构建与测算了 2006～2019 年中国 283 个地级市的人民币实际有效汇率以及二氧化碳排放强度，利用面板数据考察了人民币实际有效汇率波动对地区二氧化碳排放强度的影响程度以及作用机制。通过分析，我们得出以下结论：（1）人民币实际有效汇率升值具有明显的"碳减排效应"。其中汇率每升值 1%，地区的二氧化碳排放强度平均下降 0.463 吨/万元；（2）人民币升值的"碳减排效应"在东部地区、沿海地区、城镇化水平较高地区、信息公开地区更为明显；（3）人民币实际有效汇率升值通过提升地区的研发创新能力、抑制其外贸和外资水平、促进产业结构优化升级、改善收入不平等现象，从而有效降低二氧化碳排放强度。

基于上述研究发现，本文提出以下几点建议：

第一、深入贯彻新发展理念，积极发挥科技创新在碳减排过程中的作用。面对汇率波动时，企业应进一步提高研发创新能力以及对外部先进技术的选择、消化和吸收能力，不仅要关注生产技术创新，而且也要加强绿色技术创新，加快绿色转型的步伐，积极落实"双碳"目标。同时，各级政府应加大对绿色创新项目的投资和扶持力度，尤其要重点关注中西部能源产区的绿色创新行为，推进城市化进程，促使绿色低碳发展实现区域平衡。

第二，在坚持高水平对外开放的同时，我国政府应该积极发挥汇率杠杆的经济调节功能。一方面，利用汇率政策来改善对外贸易结构，使出口产业由资源密集型产业向技术密集型产业进行转变，促进加工贸易转型升级，改善行业间收入不平等现象；另一方面，利用汇率政策来提高外商投资的质量，改变之前"重数量、轻质量"的引资模式，避免成为发达国家的"污染天堂"，并且对低能耗、高技术、高效益的"绿色"外资进行筛选，引导其流向节能环保领域，在推动经济增长的同时进一步降低地区二氧化碳排放强度。

第三，在制定碳减排的相关政策时，政府部门应带着全国一盘棋实现"双碳"目标的思路，充分考虑中国内部的地区差异，从地区实际出发，制定差异化发展战略。对于东部地区来说，要想实现"双碳"目标，最主要的是提升能源利用效率、降低单位 GDP 能耗，加快转变经济发展模式、优化产业结构，加大研发力度、开发新能源新技术，使其在节能减排方面发挥示范作用；对于中西部地区而言，则应减轻重工业比重、优化能源消费结构、大力发展清洁能源，协同推进经济发展与环境保护。

# 参考文献

[1] 曹伟. 货币升值、经济结构转型与构建新发展格局研究[J].经济学家, 2023, 291(03):79-88.
[2] 曹伟, 金朝辉, 邓贵川, 万谍. 人民币汇率变动、资源转移与产业结构升级[J].财贸经济, 2023, 44(03):86-102.
[3] 陈诗一, 陈登科. 雾霾污染、政府治理与经济高质量发展[J].经济研究, 2018, 53(02):20-34.
[4] 干杏娣, 陈锐. 人民币升值、进出口贸易和中国产业结构升级[J].世界经济研究, 2014, 247(09):16-22.
[5] 龚利, 屠红洲, 龚存. 基于 STIRPAT 模型的能源消费碳排放的影响因素研究——以长三角地区为例[J].工业技术经济, 2018, 37(08):95-102.
[6] 韩峰, 谢锐. 生产性服务业集聚降低碳排放了吗?——对我国地级及以上城市面板数据的空间计量分析[J].数量经济技术经济研究, 2017, 34(03):40-58.
[7] 何枫, 徐桂林. FDI 与我国城乡居民收入差距之间是否存在倒 U 形关系[J].国际贸易问题, 2009, 323（11）:89-96.
[8] 井波, 倪子怡, 赵丽瑶, 刘凯. 城乡收入差距加剧还是抑制了大气污染?[J].中国人口·资源与环境, 2021, 31(10):130-138.
[9] 康志勇. 资本品、中间品进口对中国企业研发行为的影响："促进"抑或"抑制"[J].财贸研究, 2015, 26(03):61-68.
[10] 黎振强, 周秋阳. 产业结构升级是否有助于促进碳减排——基于长江经济带地区的实证研究[J].生态经济, 2021, 37(08):34-40.
[11] 李宏彬, 马弘, 熊艳艳, 徐嫄. 人民币汇率对企业进出口贸易的影响——来自中国企业的实证研究[J].金融研究, 2011, 368(02):1-16.
[12] 李向前, 袁雅静, 贺卓昇. 人民币汇率变动对产业结构的影响研究[J].西南民族大学学报(人文社科版), 2019, 40(08):120-126.
[13] 李颖, 高建刚. 人民币汇率变动、城乡收入差距与居民消费[J].广东财经大学学报, 2016, 31(03):43-56.
[14] 路正南, 罗雨森. 空间溢出、双向 FDI 与二氧化碳排放强度[J].技术经济, 2021, 40(06):102-111.
[15] 马丹, 陈紫露. 人民币实际汇率对收入不平等的非对称影响研究:基于 NARDL 模型的实证检验[J].世界经济研究, 2020, 312(02):47-58.
[16] 毛日昇. 人民币汇率与中国 FDI 流入：基于双边的视角[J].经济与管理评论, 2015, 31(04):93-105.
[17] 任桐瑜, 李杰. 人民币汇率波动与收入不平等：影响渠道及应对[J].金融发展研究, 2021, 469(01):3-13.
[18] 孙丽文, 李翼凡, 任相伟. 产业结构升级、技术创新与碳排放——一个有调节的中介模型[J].技术经济, 2020, 39(06):1-9.
[19] 孙少勤, 左香草. 汇率变动、进口中间品质量与我国全要素生产率[J].东南大学学报(哲学社会科学版), 2020, 22(01):71-80.
[20] 王少平, 欧阳志刚. 中国城乡收入差距对实际经济增长的阈值效应[J].中国社会科学, 2008, 170(02):54-66.
[21] 王松奇, 徐虔. 人民币汇率变动对产业结构影响的实证研究[J].经济理论与经济管理, 2015, 300(12):5-18.
[22] 魏巍贤, 杨芳. 技术进步对中国二氧化碳排放的影响[J].统计研究, 2010, 27(07):36-44.
[23] 谢建国, 丁方. 外商直接投资与中国的收入不平等——一个基于中国东部省区面板数据的研究[J].世界经济研究, 2011, 207(05):57-63.
[24] 鄢哲明, 杜克锐, 杨志明. 碳价格政策的减排机理——对技术创新传导渠道的再检验[J].环境经济研究, 2017, 2(03):6-21.
[25] 杨传明. 新旧常态下不同来源技术进步对中国产业碳强度的影响[J].技术经济, 2019, 38(02):112-120.
[26] 于燕, 杨志远. 行业 R&D 强度视角下中国进口贸易的技术溢出效应[J].世界经济研究, 2014, 242(04):44-50.
[27] 余淼杰, 王雅琦. 人民币汇率变动与企业出口产品决策[J].金融研究, 2015, 148(04):19-33.
[28] 余泳泽, 王岳龙, 李启航. 财政自主权、财政支出结构与全要素生产率——来自 230 个地级市的检验[J].金融研究, 2020, 475(01):28-46.
[29] 张兵兵, 徐康宁, 陈庭强. 技术进步对二氧化碳排放强度的影响研究[J].资源科学, 2014, 36(03):567-576.
[30] 张云辉, 郝时雨. 收入差距与经济集聚对碳排放影响的时空分析[J].软科学, 2022, 36(03):62-67.
[31] 赵冰, 钟霖. 人民币升值对我国进出口贸易的影响[J].统计与决策, 2010, 324(24):110-111.
[32] 钟超, 刘宇, 汪明月, 史巧玲. 中国碳强度减排目标实现的路径及可行性研究[J].中国人口·资源与环境, 2018, 28(10):18-26.
[33] Daron, A., Philippe, A., Leonardo, B., and David, H. The Environment and Directed Technical Change[J]. American Economic Review, 2012, 102(01): 131-166.

[34] Dogan, E., Seker, F., and Bulbul S. Investigating the Impacts of Energy Consumption, Real GDP, Tourism and Trade on CO2 Emissions by Accounting for Cross−Sectional Dependence: A panel Study of OECD Countries[J]. Current Issues in Tourism, 2017,20(16): 1701-1719.
[35] Ekholm, K., Moxnes, A., and Ulltveit−Moe, K. H. Manufacturing Restructuring and the Role of Real Exchange Rate Shocks[J]. Journal of International Economics, 2012, 86(01): 101-117.
[36] Esty, D. C., and Dua, A. Sustaining the Asia Pacific Miracle: Environmental Protection and Economic Integration[M]. Washington, D. C: Institute for International Economics,1997:57-60.
[37] Esty, D., and Geradin, D. Market Access, Competitiveness, and Harmonization: Environmental Protection in Regional Trade Agreements[J]. Harvard Environmental Law Review, 2004, 21(02): 265-336.
[38] Glaeser, E. L., and Kahn, M. E. The Greenness of Cities: Carbon Dioxide Emissions and Urban Development[J]. Journal of Urban Economics, 2010, 67(03): 404-418.
[39] Kim, J., and Wilson, J. D. Capital Mobility and Environmental Standards: Racing to the Bottom with Multiple Tax Instruments[J]. Japan and the World Economy, 1997, 9(04): 537-551.
[40] Li, H., Ma, H., and Xu, Y. How Do Exchange Rate Movements Affect Chinese Exports? — A Firm−level Investigation[J]. Journal of International Economics, 2015, 97(01): 148-161.
[41] Liang, F. Does Foreign Direct Investment Harm the Host Country's Environment? Evidence from China[J]. Current Topics in Management,2014, 17(03): 105-121.
[42] Omri, A., Nguyen, D. K., and Rault, C. Causal Interactions Between CO2 Emissions, FDI, and Economic Growth: Evidence from Dynamic Simultaneous—Equation Models[J]. Economic Modelling, 2014, 42: 382-389.
[43] Sapkota, P., and Bastola, U. Foreign Direct Investment, Income, and Environmental Pollution in Developing Countries: Panel Data Analysis of Latin America[J]. Energy Economics, 2017, 64: 206-212.
[44] Scruggs, L. A. Political and Economic Inequality and the Environment[J]. Ecological Economics, 1998, 26(03): 259-275.
[45] Shah, M. H., Ullah, I., Salem, S., Ashfaq, S., Rehman, A., Zeeshan, M., and Fareed, Z. Exchange Rate Dynamics, Energy Consumption, and Sustainable Environment in Pakistan: New Evidence From Nonlinear ARDL Cointegration[J]. Frontiers in Environmental Science, 2022, 9: 814666.
[46] Sun, C., Zhang, F. and Xu, M. Investigation of Pollution Haven Hypothesis for China: An ARDL Approach with Breakpoint Unit Root Tests[J]. Journal of Cleaner Production, 2017, 161(10): 153-164.
[47] Zhang, Y., and Zhang S. The Impacts of GDP, Trade Structure, Exchange Rate and FDI Inflows on China's Carbon Emissions[J]. Energy Policy, 2018, 120(09): 347-353.